\documentstyle[prl,aps,twocolumn,epsfig]{revtex}
\begin{document}

\draft
\title{Flux Lattice Melting and Lowest Landau Level Fluctuations} 
\twocolumn[ 
\hsize\textwidth\columnwidth\hsize\csname@twocolumnfalse\endcsname 

\author{Stephen W.~Pierson$^1$ and Oriol T.~Valls$^2$}    
\address{$^1$Department of Physics, Worcester Polytechnic
Institute (WPI), Worcester,    MA 01609-2280}    
\address{$^2$School of Physics and Astronomy and
Minnesota Supercomputer Institute \\  University of Minnesota, Minneapolis, 
MN 55455-0149}  

\date{November 24, 1997}
\maketitle

\begin{abstract} 
We discuss the influence of lowest Landau level (LLL) fluctuations near 
$H_{c2}(T)$ on flux lattice melting in YBa$_2$Cu$_3$O$_{7-\delta}$ (YBCO). 
We show that the specific heat step of the flux lattice melting 
transition in 
YBCO single crystals can be attributed largely to the degrees of freedom
associated with LLL fluctuations. 
These degrees of freedom
have already been shown to account for most of the latent heat. 
We also show that these results are a consequence of the correspondence
between flux lattice
melting and the onset of LLL fluctuations. 
\end{abstract} 
]
\narrowtext
Recent high-quality specific heat 
measurements\cite{junod96b,junod97,schilling97} on YBa$_2$Cu$_3$O$_{7-\delta}$ (YBCO) 
single crystals have uncovered features
that leave little doubt that the flux lattice melts via a first order 
phase transition into a new state of matter called a 
vortex liquid.\cite{bishop96} 
The sharp spikes observed in these measurements at fields of
up to 16 Tesla (T) reinforce the 
previous indications of flux lattice melting (FLM) in earlier specific heat 
measurements\cite{schilling96,junod96,welp96b} and resistivity and 
magnetization experiments.\cite{safar92,welp96,liang96} There are many 
theoretical\cite{nelson85,nelson89,houghton89,sengupta91,herbut95,yu98} 
treatments and numerical 
simulations\cite{teitel93,carneiro95,nguyen96,sasik95,hu97,blatter97} 
which treat the flux lattice melting in
both the low and high field regimes. The considerable breadth of
the theoretical approaches brought to bear on the question has produced
a considerable number of insights, but no clear
overall consensus has yet evolved as to the origin of the 
features that the experiments have revealed. 

A prominent feature of the specific heat results
is the spikes associated with the heat of melting.
Along with these spikes, steps were reported  
in fields up to 9T in Ref.~\onlinecite{schilling97}, 
with a larger specific heat on the vortex liquid side of the transition. 
Such steps were also observed in
Refs.~\onlinecite{junod97,schilling96,junod96}. 
Schilling {\it et al.}\cite{schilling97} found that they 
were unable to explain the steps in terms of the 
Abrikosov ratio, the effective 
Debye temperature, the number of vortices or the
vortex degrees of freedom.
Indeed, other authors have also shown
that degrees of freedom not associated with the vortices
contribute a  significant amount to the entropy jump (i.e. the
specific heat spike) at the FLM 
transition. For example, Hu and 
MacDonald\cite{hu97} found in their Monte Carlo study that 90\% of the latent 
heat at the FLM transition comes primarily from ``the change in 
entropy content at microscopic length scales associated with a change 
in the magnitude of the superconducting order parameter and not from 
changes in the entropy content of vortex configurations."\cite{hu97} 
Alternatively, it was suggested by
Volovik\cite{volovik97} that some of the excess
entropy could be attributed to 
``electronic''
degrees of freedom in the vortex background, that is, to
quasiparticle excitations 
close to the gap nodes of a d-wave 
superconductor. In any case, it appears that the vortices are
not the leading contributors to the latent heat.

In this paper, we address the question of the specific heat
{\it step} which is observed in connection with the spike.
We show that, for fields larger than 2-3 T, the entropy 
from lowest-Landau-level (LLL) fluctuations provides a significant,
if not leading,
contribution to the steps observed in the specific 
heat at the FLM transition. Such an explanation for the 
steps implies a deeper connection between FLM and LLL 
fluctuations: namely, that flux-lattice melting corresponds 
with the onset of LLL fluctuations. 
Evidence for such a correspondence has been presented 
before,\cite{scfun} but the arguments for this idea will be reinforced
and made more persuasive here.
We develop our argument along the following lines:
First, we will show that
an analysis of the specific data of Ref.~\onlinecite{schilling97} 
in terms of the analytical LLL expressions of 
Refs.~\onlinecite{zlatko92,zlatko94} 
indicates that a good portion of the ``step'' near the spike 
can be attributed to the onset of LLL fluctuations. 
Second, a comparison of the spike positions with LLL 
predictions\cite{herbut95,herbut94,hikami91} emphasizes the correspondence of 
flux-lattice melting with the onset of LLL fluctuations.

It is helpful to briefly review fluctuations in superconductors 
and the terminologies commonly employed to describe them. 
We are particularly concerned here with LLL fluctuations. 
In conventional bulk 
superconductors, there is a phase transition at $H_{c2}(T)$. 
Whether or not fluctuations are important is determined by the 
Ginzburg criterion.\cite{ginzburg60} If fluctuations are negligible, the 
signature of the transition in specific heat measurements as a function
of temperature consists of
ramps with the mean-field discontinuity
at the transition.\cite{hake64} If fluctuations 
are significant, they contribute a ``bump'' on top of this mean-field ramp 
and there is no sharp discontinuity.\cite{farrant75} 
These fluctuations might be generally denoted as $H_{c2}$ fluctuations.
At higher fields 
(larger than about 1-2T as specified below), the system can be treated
within a Ginzburg-Landau, 
lowest-Landau-level formalism. In this case, these fluctuations 
are called LLL fluctuations. Fluctuations can contribute to the entropy 
through
microscopic order parameter 
amplitude fluctuations as well as vortex position fluctuations,
represented by zeroes of the order parameter. The
former are those found responsible in the simulations of Hu and
MacDonald\cite{hu97} for most of the entropy change.

Analytical expressions for the specific heat of two-dimensional 
(2D), layered, and three-dimensional (3D) superconductors have 
been derived\cite{zlatko92,zlatko94} through the use
of a Ginzburg-Landau LLL approach. 
We will use these theoretical results to fit
the YBCO, $H\parallel c$ data (for field values 3T-9T) of 
Ref.~\onlinecite{schilling97}. We will use the 
3D expressions,\cite{zlatko94} since YBCO is
relatively isotropic when compared to the Bismuth-, Mercury-, or 
Thallium- based copper oxides.
This 3D function is written in detail as Eq.~(2) of Ref.~\onlinecite{scfun} 
and it provides a good fit to the data for fields $H\gtrsim 2 T$ for 
YBCO type materials.\cite{zlatko94,pierson95com,pierson96} 
In the data we consider from Ref.~\onlinecite{schilling97}, 
the zero-field data was subtracted off. Since 3D LLL expressions
do not apply to zero field, we will approximate the theoretical
zero-field contribution by a mean-field 
expression. This is valid for the temperature ranges that we are 
investigating since the fluctuations in zero field are negligible 
here.
\begin{figure}
\centerline{
\epsfig{file=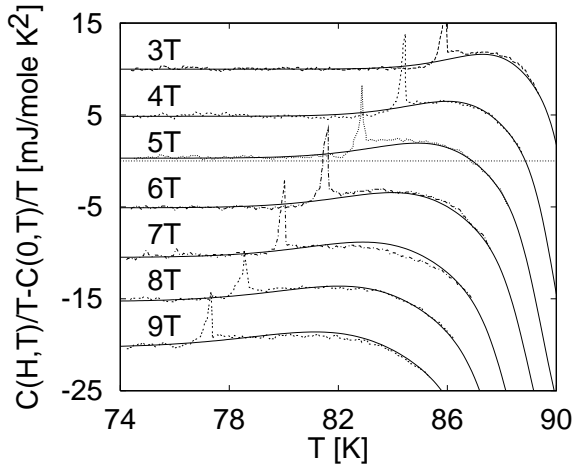,width=3.4in}}
\protect\caption{The $H\parallel c$ YBCO specific heat data of 
Ref.~\protect\onlinecite{schilling97} along with fits,
discussed in the text, to the 3D specific heat 
function of Ref.~\protect\onlinecite{zlatko94}.}
\label{schillingfit} 
\end{figure}
In Fig.~\ref{schillingfit}, the best fit to Schilling {\it et al.}'s 
data\cite{schilling97} is shown. The fits are quite 
satisfactory, except of course in the regions of the spikes, 
and the curves provide a smooth crossover from the vortex solid 
phase to the vortex liquid phase. Thus, we can see
that a major portion of the ``step'' near the spike can 
be attributed to the onset of of $H_{c2}$ or LLL fluctuations.  
The remaining 
part of the ``steps" might be explained, at least to some extent, 
in terms of the thermodynamic-equilibrium properties of the first 
order vortex lattice phase transition discussed in 
Ref.~\onlinecite{schilling97} and perhaps also in part
by the quasiparticle excitations.\cite{volovik97}
The fitting parameters used 
in Fig.~\ref{schillingfit}
are the c-axis coherence length $\xi_c=7.218\AA$, the ratio 
of the slope of $H_{c2}(T)$ to the Ginzburg-Landau parameter 
$\kappa$: $H_{c2}'/\kappa=3.40\times 10^{-2} T/K$, the mean-field 
transition temperatures 
$T_c(H)=91.33$, $90.86$, $90.50$, $90.25$, $89.8$, $89.51$, 
$89.2K$ for $H=3T-9T$, and the parameters $Q=8.29$, $K=-1.44$, and $M=5.21$ of 
Ref.~\onlinecite{zlatko94}. The values of all parameters are 
reasonable. For the YBCO materials typical values are $H_{c2}'=1.8 T/K$, 
$\kappa=52$, and $\xi_c\simeq 3\AA$. The values of $T_c(H)$ do 
produce an $H_{c2}'$ which is large. The most likely reason for 
this, in our opinion, is the fact that the function does not account 
for 3D/2D crossover. An extensive discussion  of this point is in 
Ref.~\onlinecite{scfun}.

\begin{figure}
\centerline{
\epsfig{file=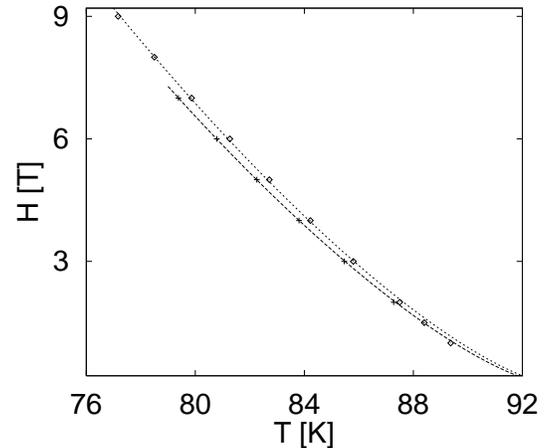,width=3.4in}}
\protect\caption{The positions of the FLM features of 
Ref.~\protect\onlinecite{schilling97} (triangles) and 
Ref.~\protect\onlinecite{junod96} 
(plus signs) plotted in $H$-$T$ space along with their 
respective fits (see text) to 
Eq.~(\protect\ref{HvT}).}
\label{HvTfig} 
\end{figure}

We turn now to the second point. Our evidence that the LLL 
fluctuations contribute more to the specific heat on the vortex 
liquid side of the transition than the vortex lattice side implies
that FLM coincides with the onset of LLL fluctuations. To develop 
this correspondence, we turn to
the 3D LLL calculation prediction that the melting temperature 
$T_M(H)$ should occur at a fixed value of the reduced temperature
$y\equiv(T_M-T_c(H))/(T_M H)^{2/3}={\rm constant}$. Herbut and 
Te\v sanovi\' c\cite{herbut95,herbut94} have calculated the value of the 
scaling constant using density functional theory finding,
\begin{eqnarray} 
(T_M-&&T_c(H))/(T_M H)^{2/3}\nonumber\\
&&=(32\pi^2\sqrt{10.5T_{c0}} \kappa^2 \xi_{ab}^2 
k_B/[\phi_0^2H_{c2}'\xi_c])^{2/3}.
\label{HvT}
\end{eqnarray}
(A similar value for the constant was calculated by 
Hikami {\it et al.}\cite{hikami91}
using perturbative series expansions.)
Here we compare the experimental features in the specific 
heat data, which mark the melting, to Eq.~(\ref{HvT}). We have done this
for two sets of data, as shown in Fig.~\ref{HvTfig}. 
The spikes of Ref.~\onlinecite{schilling97} are
denoted by the triangles. The dashed line through them
is a two-parameter fit of this theory to the 
positions of the spikes, using $H_{c2}'=1.8 T/K$, 
and a linear 
$T_c(H)$. We find a mean-field transition temperature 
$T_{c0}=93.07 K$ and that the constant in the above equation 
equals to $0.1379 K^{1/3}/T^{2/3}$. The standard 
deviation is $0.05$. The value of the constant can be 
calculated from the right hand side of the 
equation using 
$H_{c2}'=1.8 T/K$, $\kappa=52$, $\xi_c=3\AA$, and $\xi_{ab}=17.8\AA$.
These are all within reasonable range. 
We have done a similar analysis for the features observed in 
Ref.~\onlinecite{junod96} (plus signs in the Figure)
associated with FLM. This fit is also shown in Fig.~\ref{HvTfig}.
We find
$T_{c0}=92.92 K$ and const=$0.1427 K^{1/3}/T^{2/3}$  which 
would correspond to $\xi_{ab} = 
18.26 \AA$. As one can see in Fig.~\ref{HvTfig}, the fits 
to both sets of data are very good. There is somewhat more 
deviation at the lower fields ($H\sim 1-2T$),
which is reasonable since that is where the LLL 
approximation is expected to break down. That the values of the
fitting parameters to data
from two separate YBCO samples are reasonable and nearly the 
same reinforces the idea that FLM corresponds with the onset 
of LLL fluctuations. 

Evidence for the correspondence of FLM with the onset 
of LLL fluctuations has been previously found\cite{scfun} using the 
approach of Roulin {\it et al.}\cite{junod96} These authors 
identified the peaks in the
differential $C(H+\delta H,T)-C(H,T)$ with the flux
lattice melting temperature. In Ref.~\onlinecite{scfun}, it was
shown that the peaks in the differential could be partially
accounted for by the onset of LLL fluctuations. In particular,
peaks in the differential of ``theoretical'' data, 
generated using the functions of Ref.~\onlinecite{zlatko94}, 
were used to identify the temperatures of the onset of 
LLL fluctuations which were 
then shown to correspond with flux lattice melting 
temperatures found in experiments.\cite{junod96,welp96}

Associating FLM with the onset of LLL fluctuations may have 
escaped previous researchers because one does not expect such 
fluctuations to extend to temperatures so much lower than 
$T_{c2}(H)$. Yet, simple estimates using the Ginzburg 
number do reveal that in zero field fluctuations can 
become significant at temperatures on the order of five 
Kelvin\cite{herbut95} (even more in the presence of a 
magnetic field) below this temperature. Furthermore 
such a correspondence is not inconsistent with the 
behavior of conventional superconductors where FLM 
and $H_{c2}(T)$ are indistinguishable since the 
Ginzburg criterion is several orders of magnitude 
smaller than in high-temperature superconducting 
materials. 

The statement that FLM corresponds with the onset 
of LLL fluctuations could be recast in terms of a field-dependent Ginzburg
number $Gi(H)$. One can simply say that FLM is determined by
$Gi(H)$. The usual Ginzburg number is only defined in zero field. 
The field-dependent Ginzburg criterion says that fluctuations become
important when $(T-T_c(H))/T_c(H)\simeq Gi(H)$. Since we have found
evidence that the fluctuations become important at the FLM temperature,
$Gi(H)$ could then be introduced from Eq.~(\ref{HvT}). One finds,
\begin{equation} 
Gi(H)\simeq H^{2/3}(32\pi^2\sqrt{10.5} \kappa^2 \xi_{ab}^2 
k_B/[\phi_0^2H_{c2}'\xi_c])^{2/3}.
\label{GiB}
\end{equation}
This value is seven times larger than the estimate 
given by Blatter {\it et al.}\cite{blatter94} 

In summary, we have shown that the 
specific heat steps observed at the 
FLM transition in the high-quality specific heat measurements 
of Schilling {\it et al.}\cite{schilling97} originate mainly
in the  entropy associated with lowest Landau level fluctuations.
Thus, the step appears to be amenable to the same
explanations as those for the large entropy jumps given in 
Ref.~\onlinecite{hu97}. We have further shown 
that the FLM features correspond with the onset of LLL 
fluctuations and have derived an expression for the field-dependent
Ginzburg number that applies to fields where the LLL approximation is 
valid. We speculate that at lower fields, FLM corresponds with 
the onset of what we have called $H_{c2}$ fluctuations.

We gratefully acknowledge conversations with Igor Herbut, 
Z.~Te\v sanovi\' c, R.~\v S\'a\v sik, Jun Hu, N.~E.~Phillips, and A.~Schilling. 
Acknowlegement is made by SWP to the donors of The Petroleum 
Research Fund, 
administered by the ACS, for support of this research.

\end{document}